\documentclass{emulateapj}   
\usepackage[cp1251]{inputenc}
\usepackage[english]{babel}
\usepackage[dvips]{color} 
\usepackage{natbib}
\usepackage{latexsym,epsfig,graphics,graphicx,amsmath,ulem}
\input{epsf}

\def\be{\begin{equation}}
\def\ee{\end{equation}}
\def\ba{\begin{eqnarray}}
\def\ea{\end{eqnarray}}

\def\12{{1\over 2}}

\def\msun{M_\odot}

\def\etal{{\it et~al.}}
\def\ltsima{$\; \buildrel < \over \sim \;$}
\def\simlt{\lower.5ex\hbox{\ltsima}}
\def\gtsima{$\; \buildrel > \over \sim \;$}
\def\simgt{\lower.5ex\hbox{\gtsima}}

\begin{document}

\title{{HI absorption from the epoch of reionization and primordial magnetic fields}}
\author{Evgenii O. Vasiliev\altaffilmark{1} ,  Shiv K. Sethi\altaffilmark{2}}
\altaffiltext{1}{Southern Federal University, Rostov on Don 344090, Russia}
\altaffiltext{2}{Raman Research Institute, Bangalore 560080, India}

\begin{abstract}
We study the impact of primordial magnetic fields on the HI absorption
 from the Epoch of Reionization. The presence of these fields result
in two distinct effects: (a) the heating of the haloes from   the decay
of magnetic fields owing to  ambipolar diffusion, and (b) an increase
 in the number of haloes owing to additional matter fluctuations induced by
magnetic fields. We analyse both these effects and show that the latter
 is potentially observable because the number of haloes along of line of sight 
can increase by many orders of magnitude. While this effect is not strongly
dependent on the magnetic field strength in the range
 $0.3\hbox{--}0.6$~nG, it is extremely sensitive to the magnetic field power 
 spectral index for the near scale-free models.  Therefore the 
detection of such  absorption features
could be a sensitive probe of the  primordial magnetic field and 
its power spectrum. We discuss the detectability of these features with 
the ongoing and future radio interferometers. In particular, we show that 
LOFAR might be able to detect these absorption features at $z \simeq 10$
in less than 10~hrs of integration if the flux of the background source is
400~mJy. 
\end{abstract}

\section{Introduction}

Future observations of the 21~cm line from neutral hydrogen from the redshift range 10--50 will be an 
important source of information about, firstly, the  study of global 21 cm signal after the recombination, 
which depends on the evolution of the thermal state of the universe { \citep[e.g.][]{2010PhRvD..82b3006P,2006PhR...433..181F,sethiHI,2004ApJ...608..611G,shaver99},}
and, secondly, the formation of first stellar objects and reionization topology through the statistical 
properties of the 21 cm signal \citep[e.g.,][]{tozzi,iliev02}. These observations can also help to put
constraints on the relics of the inflation, such as super-heavy dark matter particles, which can produce 
ultra-high energy cosmic rays \citep{sh04,vs06,sh07}, annihilation and decays of dark matter particles
\citep{furl06,sh07,chuzhoy07,nusser07,leptonddm,silk21ddm,natarajan} and primordial magnetic fields 
\citep{sethi05,tashiro06, 2009JCAP...11..021S}. 

The first two relics lead to extra ionization and  heating of the gas both in the background 
\citep[e.g.,][]{sh07, chuzhoy07,nusser07} and in minihalos \citep{vs13}. The impact of primordial 
magnetic fields  is more complicated. The decay of these fields in the post recombination era 
alters  the thermal and ionization evolution of the gas  which influences  the properties of the 21 cm
global signal \citep{sethiHI,sethi05,tashiro06,schleicher09}. Also additional ionization and heating  
stimulate a growth of the molecular hydrogen fraction \citep{sethiH2}, that can influence on the 
formation of first luminous objects. 

More significant effect caused by primordial magnetic fields originate from the additional density 
fluctuations induced  by these fields \citep{wasserman78,kim96,subramanian98,gopal03}. These effects
act to both cut the number of haloes below the magnetic Jeans' scale and increase the number of 
haloes at larger scales owing to additional density perturbations  \citep[e.g.][]{sethi05}.

Unlike super-heavy dark particles and decaying dark matter, which have no observational counterpart in 
the local Universe, { primordial magnetic field of nano-Gauss strength  may be the progenitor of the
large-scale magnetic fields observed in  galaxies and clusters of galaxies with the coherence lengths up 
to  $\simeq$ 10\hbox{--}100 kpc \citep[for details of observations
and theoretical models  see e.g.][]{2012SSRv..166..215B,Widrow02}. Even though these fields play 
an important role in various astrophysical processes,  little  is known about the origin of large-scale 
cosmic magnetic fields  and their role in the evolutionary history of the universe. These fields could 
have originated from  dynamo amplification of very small seed magnetic fields $\simeq 10^{-20} \, \rm G$ 
\citep[e.g.,][]{Parker79,ZRS83,2005PhR...417....1B,2013A&A...560A..87S,2012MNRAS.423.3148S}. It is also 
possible that much larger primordial magnetic fields ($\simeq 10^{-9} \, \rm G$)  were generated during 
the inflationary phase \citep{TW88,Ratra92,Widrow02,ryu12} and the large scale magnetic field observed 
at the present  are the relics of these fields. In this paper, we investigate one possible implication 
of such  primordial magnetic  fields. }

The strength of primordial magnetic field can be constrained by using a 
host of cosmological observables: the cosmic microwave background measurements, early reionization 
of the universe and HI signal from the era, cosmological gravitational lensing,  and the study of 
Lyman-$\alpha$ clouds, etc. From the combination of WMAP and other CMBR data an upper limit 
$B_{0}<\hbox{a few}\ \rm nG$ is obtained \citep{paoletti12,Yamazaki10,Trivedi12}. CMBR observations are 
sensitive to large scale fields. For a single power law model, significantly tighter constraints are 
possible from the impact of these fields on smaller scales, which is also of direct relevance to us in 
this paper. From  a host of constraints from  the epoch of reionization, cosmological weak lensing, and 
the  Lyman-$\alpha$ data, one can obtain  an upper limit on magnetic field strength in the range  
$ \le 0.4\hbox{--}1$~nG  \citep{2013ApJ...762...15P, 2012ApJ...748...27P,shaw2012,kahni10,kahni12}. 
Such fields  can significantly modify the mass spectrum of the halos which results in potentially 
observable changes in the HI emission signal from the epoch of reionization 
\citep{kahni12,2009JCAP...11..021S}. In this paper, we study the influence of primordial magnetic fields 
(in the range $0.3\hbox{--}0.6$~nG)  on the statistical properties of the 21~cm absorption  from the epoch 
of reionization. In principle, primordial magnetic fields can also be directly probed by  polarization of 
the 21~cm background through Zeeman splitting \citep{cooray05}.

We assume a $\Lambda$CDM cosmology with the parameters $(\Omega_0,\Omega_{\Lambda},\Omega_m, \Omega_b,h)=(1.0,0.76,0.24,0.041,0.73)$ \citep{planck,wmap}

\section{The effects of primordial  tangled magnetic fields}

We assume the primordial magnetic field to be statistically homogeneous, isotropic, and Gaussian.  
The two-point function of the tangled fields { (in comoving wave number)} can be written as:
\begin{equation}
\langle {\tilde{B}_i({\bf q})\tilde{B}^*_j({\bf k})} \rangle 
= \delta^3_{\scriptscriptstyle D}({\bf q -k}) 
\left (\delta_{ij} - k_i k_j/k^2 \right ) M(k)
\label{eq:n6}
\end{equation}
Here $M(k)$ is the magnetic field power spectrum and $k=\vert {\bf k}\vert$ is the comoving wavenumber. 
Here  we assume: $M(k) = A k^n$  and  $n$ is the magnetic field spectral
 index;  we use $n \simeq -3$
throughout this paper (for justification of this range of spectral indices  see e.g. \citep{2013ApJ...762...15P} 
and references therein). The power law is cut-off at $k_{\rm max} \simeq  200  (1 \, {\rm nG}/B_0)$ which is determined by the dissipation 
of magnetic fields in the pre-recombination era (for details see \citep{sethi05} and references therein). 
Here we refer to magnetic field strength $B_0$ as the RMS $\langle \tilde{\bf B}^2({\bf x}) \rangle$ 
filtered at $k = 1 \, \rm Mpc$. 

The presence of magnetic field results in two distinct effects. The magnetic field dissipates in the 
collapsing halo owing to ambipolar diffusion which heats the halo.  The spatial part of the heating 
rate by ambipolar diffusion can be computed (for details see \citep{sethi05}):
\begin{equation}
\langle ({\bf \nabla}_{\bf x} \times \tilde{\bf B}) \times  
\tilde{\bf B})^2 \rangle = {7 \over 3} \int dk_1 
\int dk_2 M(k_1)  M(k_2) k_1^2 k_2^4
\label{eq:n6pp}
\end{equation}

The presence of primordial magnetic fields also result in additional matter perturbations at small scales. 
The magnetic-field induced matter power spectrum $P(k) \propto k^{3n+7}$ for $n \le -1.5$ and it is sharply 
cut-off at magnetic Jeans' scale $k_J \simeq 15 (1 \,  {\rm nG}/B_0)$\citep{kim96,gopal03}. This additional power at small scales
changes the number of haloes at masses of interest $10^6 \, {\rm M_\odot}\le M \le  10^8 \rm M_\odot$. If 
the haloes in this mass range fail to form stars then they would be observable in HI absorption. 

In this paper we take into account both these effects: (a) additional heating due to ambipolar diffusion, 
and (b) the impact of magnetic-field induced small scale power on matter power spectrum. 

\section{The model of collapsing halo, HI optical depth, and the number of haloes}

\subsection{Dark matter, gas dynamics and magnetic fields}

To model the evolution of the  dark matter we follow the prescription given by \citet{ripa}. The dark matter mass, 
$M_{DM} = \Omega_{DM} M_{halo}/\Omega_M$, is assumed to be enclosed within a certain truncation radius 
$R_{tr}$, inside which the dark matter profile is a truncated isothermal sphere with a flat core of radius 
$R_{core}$. The parameter $\xi = R_{core} /R_{\rm vir}$ is taken to be 0.1 for all simulations. Such a 
description is used to mimic the evolution of a simple top-hat fluctuation \citep[e.g.][]{padma}. 
The resulting  dark matter profile has a flattened  form \citep{burkert}.

Dynamics of baryons is described by a 1D Lagrangian scheme similar to that proposed by \citet{1d}; a reasonable
convergence is found at a resolution of 1000 zones over the computational domain. Chemical and ionization 
composition includes a standard set of species: H, H$^+$, H$^-$, He, He$^+$, He$^{++}$, H$_2$, H$_2^+$, D, 
D$^+$, D$^-$, HD, HD$^+$ and $e$, with the corresponding reaction rates \citep{galli98,stancil98}. 
The energy equation includes heating owing to ambipolar diffusion and 
different  radiative losses in the primordial 
plasma: Compton cooling, recombination and bremsstrahlung radiation, collisional excitation of HI 
\citep{cen92}, H$_2$ \citep{galli98} and HD \citep{flower00,lipovka05}. Our
computation  starts at redshift 
$z = 100$. The initial parameters --- gas temperature, chemical composition and other quantities --- are 
taken from  one-zone calculations that begin at $z = 1000$ with  values  at the end of recombination: 
$T_{gas} = T_{CMB}$, $x[{\rm H}] = 0.9328, x[{\rm H^+}] = 0.0672, x[{\rm D}] = 2.3\times 10^{-5},
x[{\rm D^+}] = 1.68\times 10^{-6}$ \citep[see references and details in][Table 2]{ripa}.  

\begin{figure*}
\center
\includegraphics[width=120mm]{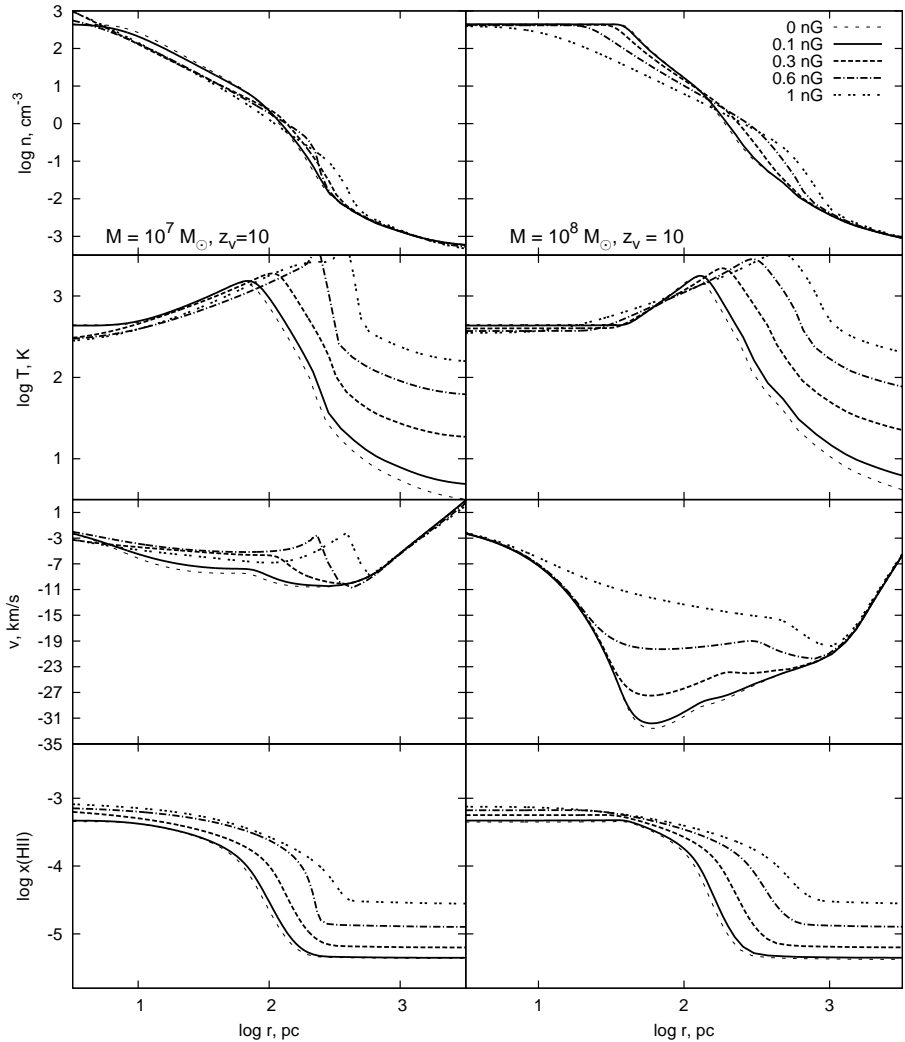}
\caption{
The  density, temperature, velocity and molecular hydrogen profiles across the collapsing halo, for 
different values of magnetic field strengths. The two panel correspond to $M = 10^7, 10^8 \, \rm 
M_\odot$, respectively for virialization redshift  $z_{\rm vir} = 10$. 
}
\label{fig:f1}
\end{figure*}

{
In the early evolution of the halo, when  it is still in the expansion phase and the 
overdensity is very small, the magnetic field strength scales as $(1+z)^2$. The initial 
condition is set during this phase. During the collapse phase of the halo, we assume the 
magnetic field strength to scale as its flux-frozen evolution with the gas density $\rho$: 
$B_0 \propto \rho^{2/3}$ \citep[e.g.][]{sethiH2,2009ApJ...703.1096S,2010ApJ...721..615S}. 
It should be noted that in  both the expansion and the collapse phase, magnetic field scales 
as $\rho^{2/3}$ with the gas density. 
}

The heating rate due to the ambipolar diffusion is computed from Eq.~(\ref{eq:n6pp}); 
during the expansion phase this rate  can be written as { \citep{2008PhRvD..78h3005S}:}
\be
L_A = {B^4 \over l_a^2} {x_{HI} \over 16\pi^2 \gamma x_e \rho^2}
\label{lax}
\ee
where $B = B_0 (1+z)^2$, $\gamma = 3\times 10^{-9}/(2m_p)$ and
$$
l_a = 1.32\times 10^{22} {B_0 \over 10^{-9}} {A\over 1+z}
$$
$A$ depends on the cosmological parameters $\Omega_m, h, \Omega_b$ and the 
magnetic field power spectrum $M(k)$. { Eq.~(\ref{lax}) can be  extended 
to the collapse phase using the flux-frozen condition discussed above. }

{
Using the method and initial conditions described above we follow the evolution
of minihaloes with masses in the range $M = 10^6 \hbox{--} 10^9 \, \rm M_\odot$; 
the  virialization redshift of these haloes is 
in the redshift range $z = 10\hbox{--}15$. In Figure~1 we show the density, temperature, 
velocity, and molecular hydrogen profiles for two haloes, including the effects of 
magnetic fields. We present the radial profiles of haloes with $M = 10^7, 10^8 \, 
\rm M_\odot$ virialized at $z_{\rm vir} = 10$ for standard recombination ($B=0$) 
and several values of initial magnetic strength: $B= 0.1,\ 0.3,\ 0.6$ and 1~nG.
The most notable feature with direct impact on results presented in this 
paper is the change in the temperature profile. As seen in the Figure, the 
presence of magnetic fields results in an increase in temperature across
the halo, which  has a direct bearing on the  strength of the observed HI profile. 
}

\subsection{21 cm optical depth and equivalent width}

The spin temperature of the HI 21 cm line is determined by atomic collisions 
and the scattering of ultraviolet 
(UV) photons \citep{field,wout}. In our calculations we used collisional coefficients from \citet{kuhlen} 
and \citet{liszt}. Magnetic fields provide the extra heating source. 
{ We do not include Lyman-$\alpha$ pumping in our study.  We note that the inclusion of Lyman-$\alpha$ 
scattering serves to couple the matter temperature with the spin temperature and plays a crucial role in 
the observability of IGM HI in emission  from the epoch of reionization \citep[e.g.][]{sethiHI,2004ApJ...608..611G}.
However, we consider only absorption from  collapsing haloes in the redshift range $10 < z< 15$ against bright radio sources. In such  haloes, the spin temperature is coupled to the matter temperature through collisions and Lyman-$\alpha$ doesn't play an important role. }

The optical depth along a line of sight at a frequency $\nu$ is: 
\ba
\label{optd} 
 \tau_\nu = {3h_p c^3 A_{10}\over 32 \pi k \nu_0^2} \int_{-\infty}^{\infty}
    dx {n_{\rm HI}(r) \over \sqrt{\pi} b^2(r) T_s(r)} 
    \\ 
      \nonumber
   \times 
  {\rm exp}\left[{-{[v(\nu) - v_l(r)]^2\over b^2(r)}}\right]
\ea
where $r^2 = (\alpha r_{\rm vir})^2 + x^2$, $\alpha=r_\perp/r_{\rm vir}$ is the dimensionless impact parameter, $v(\nu)=c(\nu-\nu_0)/\nu_0$, $v_l(r)$ is the infall velocity projected along the line of sight, 
$b^2 = 2kT_k(r)/m_p$ is the Doppler parameter.

The observed line equivalent width: $W_\nu^{obs} =W_\nu / (1+z)$, where the intrinsic equivalent 
width is 
\be 
{W_\nu} = \int_{\nu_0}^{\infty}{(1-e^{-\tau_\nu})d\nu} - 
 \int_{\nu_0}^{\infty}{(1-e^{-\tau_{\rm IGM}})d\nu}
\label{eweq}
\ee
where $\tau_{\rm IGM}$ is the optical depth of the background neutral IGM. 
Throughout this paper, we assume $\tau_{\rm IGM} = 0$

In Figure~2 we show the expected absorption profiles in the presence of magnetic fields. Additional 
heating owing to ambipolar diffusion acts to lower the expected optical depth by up to a factor of 
2--3 depending on the  strength of the field and the impact factor.

\begin{figure}
\includegraphics[width=80mm]{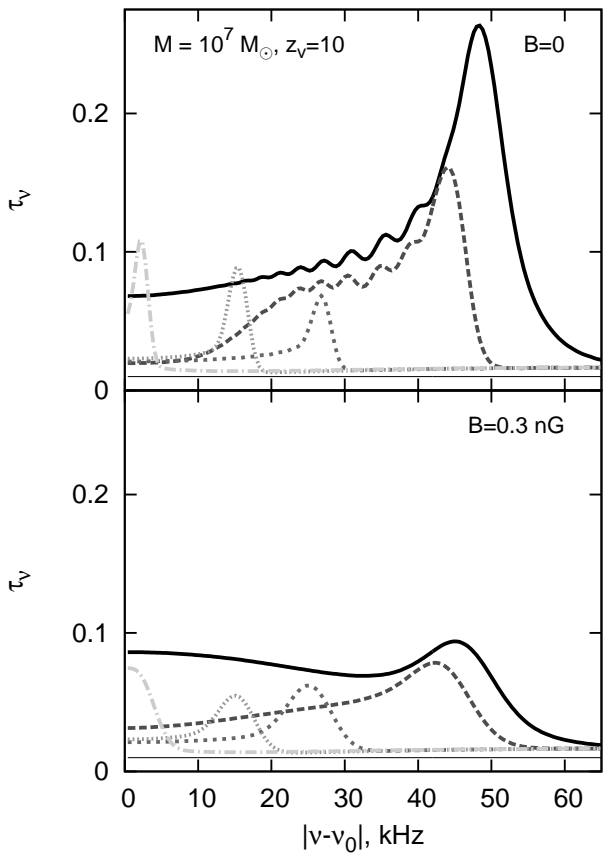}
\caption{
Optical depth at the impact parameters $0.1,\  0.3,\  1,\  1.5,\ 3r/r_{\rm vir}$ (solid, dash,
dot, short dash and dot-dash lines, correspondingly) for a halo $M=10^7\msun$  
virialized at $z_{\rm vir} = 10$ in the standard recombination model ($B=0$) and in
the presence of primordial magnetic fields with $B=0.3$~nG. The x-axis denotes
the frequency width of the signal  in the rest frame of the halo. 
}
\label{fig:f2}
\end{figure}

\subsection{Number of haloes}

We use the Press-Schechter formalism to compute the number of haloes in a given mass range at a redshift. 
This allows us to compute the number of haloes that intersect a given line of sight in a redshift range 
$z$ and $z+dz$:
\begin{equation}
N(z) = \int dM \left ({dn \over dM} \right ) \pi {r_\perp}^2  {H_0^{-1} dz \over \Omega^{1/2} (1+z)^{5/2}}
\label{eq:numz} 
\end{equation}
Here {$r_\perp$} is the impact factor and for our computation we take the maximum impact factor 
${r_\perp} = 3 r_{\rm vir}$ for any mass; $dn/dM$ is the mass function of haloes
(in $Mpc^{-3} M_\odot^{-1}$). 

The main impact of primordial magnetic fields is on the number of haloes in the mass range 
$10^6\, {\rm M_\odot}  \le M \le 10^8 \, \rm M_\odot$. Figure~3 shows the mass function of haloes 
for the models with and without the primordial magnetic fields. The number of haloes in the mass 
range of interest could be orders of magnitude larger than in the usual $\Lambda$CDM model. This 
is owing to the extra power at small scales induced by primordial magnetic fields. 
{ This extra power increases the mass dispersion  $\sigma(M,z)$ at mass scales comparable to 
the magnetic Jeans' mass; in particular such an increase might result in 1-$\sigma$ collapse of 
such haloes while the $\Lambda$CDM model only allows nearly 3-$\sigma$ overdensities  to collapse 
at $z = 10$ \citep[e.g.][]{sethi05}. In the Press-Schechter formulation, the mass function of haloes: 
$dn/dM \propto \exp[-\delta_c^2/(2\sigma^2(M,z))]$ with  $\delta_c = 1.68$; this results in a sharp 
increase in the number of haloes at small scales.}
Two other notable features of the Figure are: (a) an increase in the strength of magnetic field 
results in a decrease in the number of haloes at small mass scales and an increase for larger 
masses, this feature is caused by the increase of magnetic Jeans' length with magnetic field 
strength, and (b) the number of haloes is seen to be extremely sensitive to the magnetic field 
spectral index. For $n \simeq -3$, the mass dispersion, $\sigma(M) \propto (n+3)$. As the number 
of haloes is exponentially sensitive to the value of mass dispersion, the number of haloes sharply 
decreases as $n$ approaches $-3$. 

For $\Delta z = 0.2$ (corresponding to a frequency width $\Delta \nu \simeq 2.4$~MHz
at $z \simeq 10$) and the  $\Lambda$CDM model, $N(z) \simeq 0.1$ or one out of ten lines of 
sight is likely to intersect such a halo. For the magnetic field of strength $B_0 = 0.3$~nG  
and $n = -2.9$, we get  $N(z) \simeq 18$. 

To underline the impact of primordial magnetic fields, we define an effective  optical depth:
\ba
 \tau_{\rm eff}(z) = \int dr_\perp \int  dM \left ({dn \over dM} \right )(z) 
    \\ 
      \nonumber
   \times 
   2 \pi {r_\perp}  
       {H_0^{-1} dz \over \Omega^{1/2} (1+z)^{5/2}} \tau(M,{r_\perp}, z)
\label{taueff}
\ea
$\tau_{\rm eff}(z)$  is weighted by $\tau(M,{r_\perp}, z)$ and scales as the number of haloes so 
it captures both the decrease in optical depth owing to additional heating and the increase of number 
of haloes in the presence of magnetic field. 

\begin{figure}
\includegraphics[width=60mm,angle=270]{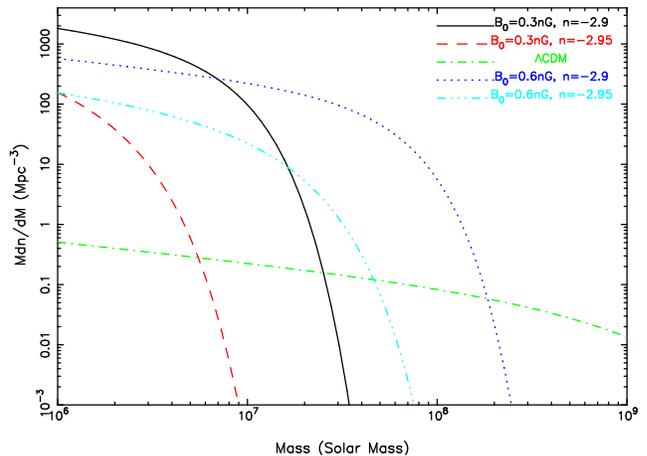}
\caption{
Comoving mass function is shown  for different models  at $z = 10$: $\Lambda$CDM model
(dot-dashed curve), solid  and dotted line ($B_0 = 0.3$~nG, $n = -2.9, -2.95$, 
respectively), dashed  and dot-dot-dot-dashed line ($B_0 = 0.6$~nG, $n = -2.9, -2.95$, respectively). 
}
\label{fig:f3}
\end{figure}

\section{Results}

\begin{figure*}
\center
\includegraphics[width=100mm,angle=270]{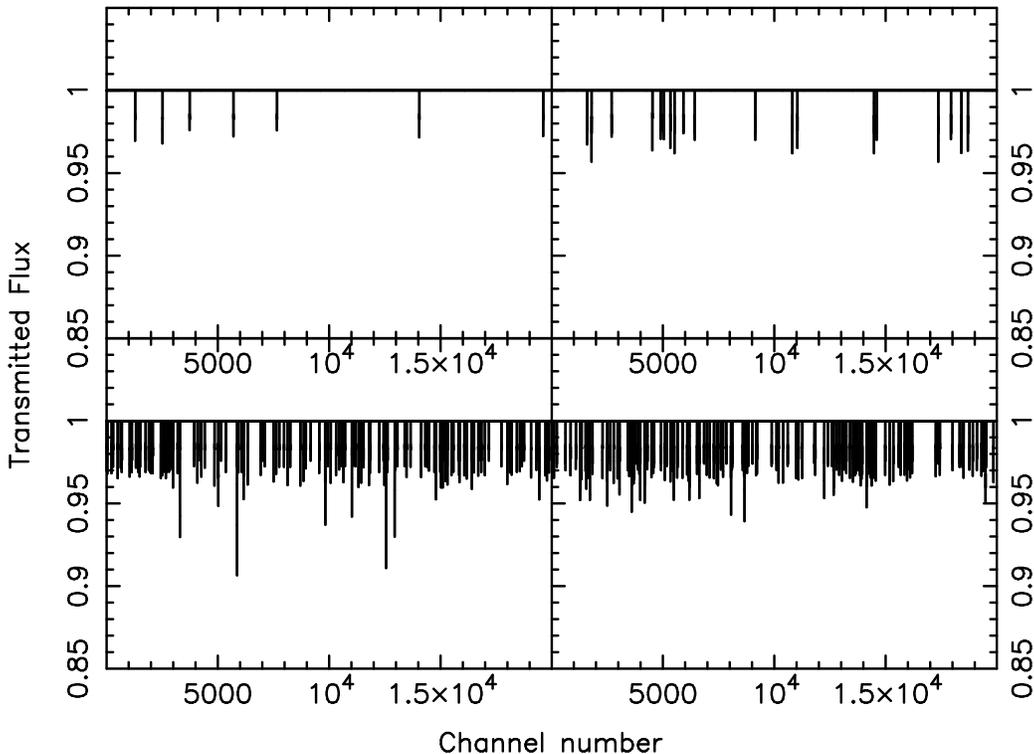}
\caption{
Simulated spectra are displayed for two values of magnetic field  $B_0= 0.3, 0.6$~nG (left and right 
panels, respectively) and for $n = -2.9, 2.95$ (lower and upper panels, respectively) at $z = 10$. The  spectra
have a channel width of 1~kHz. The total band width of each  spectrum is $\simeq 20 \, \rm MHz$.  Lower panels:  
There are 141 (7) intersecting haloes in the left lower (upper) panels and 147 (19)  in the right lower (upper) 
panels.
}
\label{fig:f4}
\end{figure*}

\begin{figure}
\center
\includegraphics[width=55mm,angle=270]{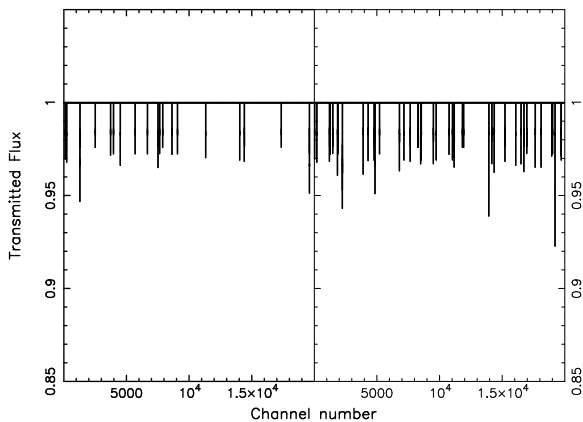}
\caption{
Simulated spectra are displayed for two values of $B_0= 0.3, 0.6$~nG (left panel) and for $n = -2.9$
(right panel) at $z = 15$. The spectra have a channel width of 1~kHz.
 The total band width of the spectra
displayed is $\simeq 20$~MHz. The number of intersecting haloes  in the left (right) panels
are 26 (45). 
}
\label{fig:f5}
\end{figure}

To illustrate the difference between the $\Lambda$CDM model and the effects of primordial magnetic 
fields, we simulate absorption spectra in the redshifted HI line. 

Eq.~(\ref{eq:numz}) gives the number of haloes in a redshift interval (or equivalently for a given 
frequency channel width). For simulating the absorption spectrum {we} choose the redshift bin  $dz$
to be such that $N(z) \ll 1$.  For each channel, a Poisson number is drawn with the average given 
by $N(z)$. Since $N(z) \ll 1$, this number is either 0 or 1. To determine the mass and impact 
factor of the intervening halo, random numbers are drawn from the mass function for the respective 
cases and probability density for the impact factor. Given the optical depth as function of mass, 
impact factor, and the frequency, this procedure allows us to simulate synthetic spectra. 

In Figure~4 we show the simulated spectra for two cases: $B_0 = 0.3$~nG and $B_0 = 0.6$~nG for 
$n = -2.9$ and $n = -2.95$ at $z = 10$. The channel width in the 
spectra is $\simeq 1\, \rm kHz$ and the total frequency coverage is 
$\simeq 20 \, \rm MHz$. 

As seen in the Figure, the number of intersecting haloes in the two cases are comparable, even though 
the number density  of haloes is larger in the former case (Figure~3). This is because  the number of 
intersecting haloes scales as $M^{1.66}dn/dm$ (Eq~\ref{eq:numz}), which puts higher weights on haloes 
of larger mass as compared to the distribution shown in Figure~3. Also this combination  picks a maximum 
at certain mass scales: for $B_0 = 0.3,\  0.6$~nG, ($n = -2.90$) the mass scales are $7 \times 10^5 \, 
\rm M_\odot$ and $6 \times 10^6 \, \rm M_\odot$, respectively (for $\Lambda$CDM model the equivalent 
mass is $2 \times 10^8 \, \rm M_\odot$). As the Figure shows, the spectrum is not particularly sensitive 
to  the strength of magnetic field, for a given value of the spectral $n$, in this range of magnetic 
field strengths. 

However, the number of haloes along a line of sight is very sensitive to the spectral index $n$.
For $n = -2.95$, the number of such sources falls by more than an order of magnitude  for both magnetic 
field strengths shown in the Figure. This means that the observed spectra could be a strong discriminator 
of the magnetic field power spectral index. 

In Figure~5 we show the simulated spectra at $z =15$. The sharp fall in the number of absorption 
features is owing to the decrease in mass dispersion at higher redshifts. 

As noted above, the $\Lambda$CDM model predicts roughly one  intersecting halo for the frequency coverage
displayed in Figure~4. Therefore, in principle, if deep absorption features seen in {Figure~4} are detected 
towards a few lines of sight, it could  be an indication of a process which produces extra matter fluctuations  
at small scales. In particular, the presence of a large number of sources along a line of sight 
(Figure~4) could be used to extract information about the magnetic field parameters.

\subsection{Detectability of HI absorption}

As discussed above, the main impact of the inclusion of magnetic field 
in our analysis is a sharp increase in the number of absorption lines (Figure~4). Each of the absorption feature has a width of $\simeq 10 \, \rm kHz$ (in the observer frame for $z \simeq 10$, Figure~2). 

This leads to at least three distinct possibilities of the detection of the signal:
\begin{itemize}
 \item[(a)] owing to 
a large number of absorption features, the probability of deep, narrow  features that arise when 
the impact parameter is small increases (Figure~2). Figure~4 shows that many such  features have
optical depths in the excess of $\tau = 0.05$; such features are detectable at frequency resolutions 
$\simeq 1 \, \rm kHz$, 
\item[(b)] at the frequency  resolution comparable to the line width we expect 
significant signal. Also there is a reasonable probability of line blending owing to 
high density of such features, 
\item[(c)] at even lower frequency resolution, one 
can hope to detect the combined effect of many lines blending into a broad 
feature \citep{ferrara11}.  
\end{itemize}

Ongoing experiments such as Murchison Widefield Array (MWA) and Low-Frequency Array (LOFAR) can in 
principle detect the absorption signal from the redshifted HI line. MWA can achieve channel widths 
$\simeq 40$~kHz and total instantaneous bandwidth of 32~MHz in the frequency range of interest 
(80--150~MHz)\footnote{\tt http://www.mwatelescope.org/index.php/telescope}. LOFAR can have a channel 
width of $\le 1$~kHz with total band width of roughly 75~MHz\footnote{\tt https://www.astron.nl/radio-observatory/astronomers/
users/technical-information/frequency-selection/
station-clocks-and-rcu}.
LOFAR spectral resolution is comparable to the spectra shown in Figure~4.

Given the angular resolution of these experiments, one of the  challenges in the detection of these 
absorption features from the epoch of reionization is to distinguish high redshift  bright continuum 
sources  from low redshift sources. MWA has an angular resolution of nearly 4' and a confusion noise 
 $\simeq 10$~mJy.  On the other hand, LOFAR presents a better prospect with an angular resolution 
$\simeq 10''$ which gives a confusion noise  $\simeq 100$~$\mu$Jy. LOFAR can reach a sensitivity of 
200~$\mu$Jy in 8~hours of integration for a bandwidth of 3.66~MHz in the frequency range 100--200~MHz. 
This corresponds to a line sensitivity (for a channel width $\Delta\nu \simeq 1$~kHz) of nearly 14~mJy. 
To detect a line feature with an optical depth $\simeq 0.05$ (Figure~4) at 5$\sigma$ level, a source 
of a few  Jansky flux density  would be needed.  Significantly longer integration time would be needed 
for fainter sources.

To investigate the possibility of detecting HI features at lower frequency
resolution, we show  equivalent widths, $W_\nu$ (Eq.~4), for spectra at two different frequency 
resolutions  in Figure~6 and~7. Even though the equivalent widths are clustered around  $W_\nu 
\le 0.2 \, \rm kHz$ in Figure~6, there are a number of features at higher equivalent widths, e.g. 
the lower, left panel of Figure~6 has two features in the range $W_\nu = 0.55\hbox{--}0.6$~kHz, 
that arise from blending of two lines. If the threshold of detection is $\tau = 0.02$, then
these features can be detected with  $\Delta\nu \simeq 30 \, \rm kHz$. 
Or at such spectral resolutions, one obtains a gain in the signal-to-noise 
of a factor of two over detecting a feature of $\tau = 0.05$ 
with  $\Delta\nu \simeq 1 \, \rm kHz$. 

In Figure~7, we display equivalent widths for spectra smoothed at $\Delta\nu \simeq 100 \, \rm kHz$. 
This allows us to assess the feasibility of detecting HI absorption features at low resolutions. 
The distribution of equivalent widths is seen to shift to larger values for the smoothed spectrum. 
This is owing to blending of densely packed spectral lines as the resolution is decreased. The most 
notable feature of the figure is that many features are seen to have equivalent widths 
$\simeq 2 \, \rm kHz$,  nearly three times the largest values for  the 
high resolution spectra.  This means that the decline in peak optical 
depth as the spectra are smoothed  is compensated by the inclusion of more  
sources. In other words, if the threshold optical depth of detection 
is 0.02, it can be detected even with a resolution of $\Delta\nu \simeq 
100 \, \rm kHz$. More specifically, using the parameters of  LOFAR, such 
features are detectable by LOFAR at 5-$\sigma$ level for 
a source flux density of $\simeq 400 \, \rm mJy$ in 8 hours of integration.

\begin{figure}
\center
\includegraphics[width=60mm,angle=270]{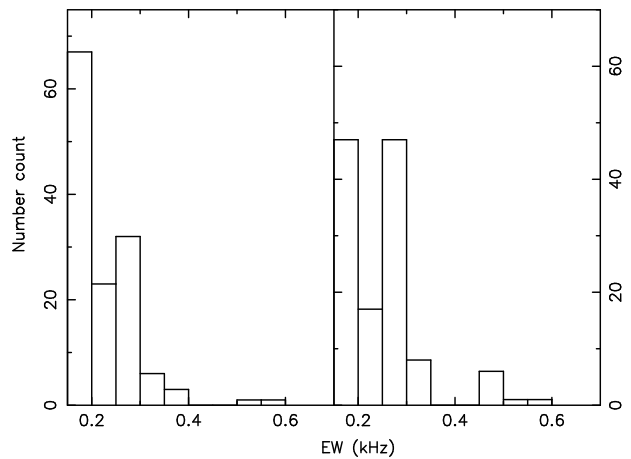}
\caption{
Histograms of equivalent width distribution for the spectra  
in Figure~4 (lower panels), 
}
\label{fig:f6}
\end{figure}

\begin{figure}
\center
\includegraphics[width=60mm,angle=270]{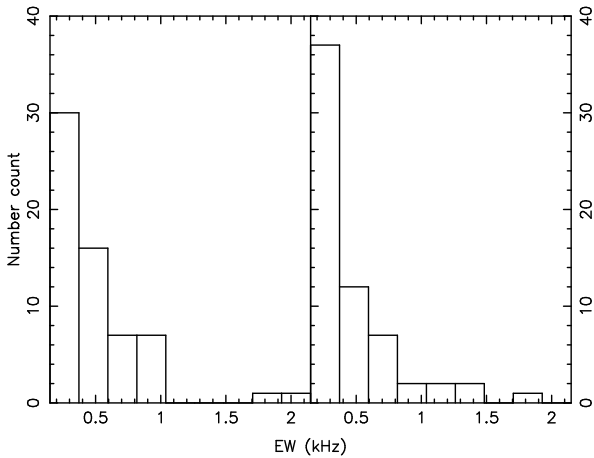}
\caption{
Histogram of equivalent width distribution for the spectra  
in Figure~4 (lower panels), smoothed by a spectral width 
of $100\, \rm kHz$. 
}
\label{fig:f7}
\end{figure}

{
As noted above, a few such sources might suffice to reveal the nature of 
magnetic field-induced extra matter power. How likely are such bright 
radio  sources at $z \simeq 10$? The radio source J0924-2201 has $z \simeq 5.2$
with a flux $F_\nu = 0.55$~Jy at 230~MHz \citep{2007AJ....133.2841C}. Such
a steep spectrum source might be brighter at $\nu \simeq 130$~MHz (redshifted 
HI frequency at $z \simeq 10$) with flux comparable to its value at 235~MHz. 
Such rare radio sources might provide a suitable setting for understanding the 
nature of density perturbations during the EoR. 
}

However, the  expected  radio flux distribution of 
sources from the epoch of reionization is highly uncertain and for 
for fainter sources the integration time might become unrealistically large, 
e.g. for a 40~mJy source, 800~hrs of integration time would be needed. 
The planned radio interferometer SKA will be ideal for detecting these 
features: its sensitivity is  projected to reach $400 \, \rm \mu$Jy in 
one minute integration in the frequency range 70--300~MHz with sub-arcsec
resolution, which will suppress the impact of confusion 
noise\footnote{\tt https://www.skatelescope.org/}.  
 
In this paper we only considered absorption from collapsed haloes which 
are rarer but give rise to large optical depths. In addition, we could have 
mildly  overdense regions (with density contrasts $\delta \simeq 1\hbox{--}10$)
in the neutral gas \citep{2012MNRAS.425.2988M}. These regions give rise to 
smaller optical depths but their effect might be detectable through 
statistical methods,  based on  distribution of radio sources during the 
epoch of reionization  \citep{2012MNRAS.425.2988M}. Magnetic fields 
alter the density field at all scales and therefore would also change  the
 statistical properties of such regions \citep[e.g.,][]{2013ApJ...762...15P}.
We hope to return to this issue in a later work.

In Table~1, we list $\tau_{\rm eff}$ for a range of models. As noted above, this 
measure captures the impact of both magnetic field heating and the change in the 
number of haloes.

\begin{deluxetable}{ccc}
\tablenum{1}
\tablecolumns{8}
\tablewidth{0pc}
\tablecaption{
The number of intersecting haloes per unit redshift and the effective optical depth 
(Eq.~(\ref{taueff})) per unit redshift are listed for a range of models with and 
without primordial magnetic fields. 
}
\tablehead{
\colhead{}&\colhead{dN(z)/dz}& \colhead{$d\tau_{\rm eff}/dz$} 
\\ 
}
\startdata
$\Lambda$CDM,  $z = 10$ & 0.25 & 0.38  \nl
$B_0=0.3$~nG, $n = -2.9$, $z = 10$  & 92  & 198  \nl
$B_0 = 0.3$~nG, $n = -2.95$,  $z = 10$  & 2  & 3.4  \nl 
$B_0 = 0.6$~nG, $n = -2.9$, $z = 10$ & 95 & 210  \nl
$B_0 = 0.6$~nG, $n = -2.95$, $z = 10$ & 10 & 22  \nl
$B_0 = 0.3$~nG, $n = -2.9$, $z = 15$ & 12 & 35  \nl
$B_0 = 0.6$~nG, $n = -2.9$, $z = 15$ & 18 & 69  \nl
\enddata
\end{deluxetable}

\subsection{Comparison between magnetic heating, dark matter decay and X-ray heating}

The evolution of the observed number and optical depths of absorption features is sensitive to the 
evolution of heating mechanisms of the IGM. Here we briefly compare the relative impact of X-ray 
heating, dark matter decay models, and the magnetic field heating. 

Using Eq.~(\ref{lax}) the heating rate due to ambipolar diffusion can be 
written as [K s$^{-1}$]: 
\be
{2 L_A \over 3 k n H(z)} \simeq 1.2 \times 10^4 \left({B_0\over 10^{-9}}\right)^2
																									 \left({1+z\over 10}\right)^{-1/2}
																									 \left({10^{-4} \over x_e}\right)
\ee

One can compare this rate to the heating rates due to X-ray and decaying dark matter.
The total normalized X-ray emissivity $\epsilon_X$ can be written as \citep{furla06}
\be
{2 \epsilon_X \over 3 k n H(z) } = 5\times 10^4 f_X \left[ {f_*\over 0.1} {df_{coll}/dz\over 0.01} 
                                  {1+z\over 10} \right], 
\label{epx}
\ee
here $df_{coll}/dz$ is the fraction of baryons collapsed to form a protogalaxy per unit redshift, 
$f_\ast$, the fraction of baryons converted into stars in a single star formation event. 

The corresponding heating rate due to DDM can be found as $K_\ast=\chi_h\epsilon_X$ \citep{vs13}
\be
{2K\over 3 k H(z) } = 8.9\times 10^4 (\xi/\xi_L) \left({1+z\over 10}\right)^{-2/3}. 
\label{kxx}
\ee

The dependence of rates Eqs~(\ref{lax}), (\ref{epx}), and~(\ref{kxx}) on the  redshift are not very 
different. So from the  global signal  it would be  difficult to separate between the  decaying dark 
matter, X-ray heating, and magnetic fields. 

In other words, the main discriminator between these models is the far larger number of haloes produced 
owing to additional density perturbations in the presence of primordial magnetic fields.

\section{Conclusions}

In the paper we studied the impact of primordial magnetic fields on HI absorption from collapsed haloes 
during the epoch of reionization. We considered haloes in the mass range $10^6-10^8~{\rm M_\odot}$ 
and the  absorption from  these haloes for $r \le 3 r_{\rm vir}$. We consider 
magnetic field strength in the range: $B_0 = 0.3\hbox{--}0.6$~nG, comparable to 
the best upper bounds on these fields from cosmological observables \citep{2013ApJ...762...15P, 2012ApJ...748...27P,shaw2012,kahni10,kahni12}. 

The presence of magnetic fields result in two separate  effects. The decay of magnetic fields in the 
collapsing haloes owing to ambipolar diffusion heats the halo which lowers the optical depth of 
absorption for intermediate masses (Figure~2). Primordial magnetic fields also generate additional 
density perturbations which result in a sharp increase in the number of haloes (Figure~3). The latter 
effect dominates the observable signature (Figure~4). 

If these haloes become star-forming they could be responsible for early  reionization and result in
distinct HI fluctuation signature from the epoch of reionization. However, if these haloes remain
HI rich then their presence will not significantly  impact the HI  emission signal from the epoch of 
reionization as their  temperatures are far larger than the CMBR temperature (Figure~1) and the HI 
emission signal is nearly independent of the spin temperature  (assuming $T_s = T_k$) in this case. 
Therefore, the most promising way to observe them would be in absorption. 

In this paper we investigate the HI absorption for $z \ge 10$. The clumpiness of HI distribution caused 
by magnetic fields might leave its signature at lower redshifts also, e.g. in the radiative transfer of 
radiation from Lyman-$\alpha$ emitting galaxies which are observed up to a redshift $z \simeq 8.5$.  

\vspace{1cm}
{
EV acknowledges partial support from the Russian Foundation for Basic Research through the grant 12-02-00365
and the "Dynasty" foundation. EV is grateful the Raman Research Institute for hospitality, where this
work was begun.
}



\end{document}